\documentclass[11pt,a4paper,DIV11]{scrartcl}
\usepackage[utf8]{inputenc}
\usepackage[T1]{fontenc}
\usepackage{lmodern}
\usepackage[british]{babel}
\usepackage{amsmath}
\usepackage{amssymb}
\usepackage{amsfonts}
\usepackage{color}
\usepackage{float}
\usepackage{stmaryrd}
\usepackage{url}
\usepackage{graphicx,psfrag}
\usepackage{placeins}
\DeclareMathOperator{\Di}{D_w}
\newcommand{\tr}{\ensuremath{\mathrm{tr}}}
\newcommand{\I}{\ensuremath{\mathrm{i}}}
\newcommand{\figg}[1]{Fig.~\ref{#1}}
\newcommand{\eqq}[1]{Eq.~\ref{#1}}
\newcommand{\secc}[1]{Sec.~\ref{#1}}
\newcommand{\tabb}[1]{Table \ref{#1}}
\providecommand{\href}[2]{#2}
\newcommand{\arxiv}[1]{[arXiv:\href{http://arxiv.org/abs/#1}{{\tt #1}}]}
\newcommand{\aetap}{\text{a--}\eta'}
\newcommand{\api}{\text{a--}\pi}
\newcommand{\afn}{\text{a--}f_0}
\newcommand{\saetap}{\text{a}\mbox{-}\eta'}
\newcommand{\sapi}{\text{a}\mbox{-}\pi}

\begin{document}
\title{The gluino-glue particle and finite size effects in supersymmetric
Yang-Mills theory
\vspace*{3mm}}
\author{G.~Bergner, T.~Berheide, G.~M\"unster, U.\,D.~\"Ozugurel, D.~Sandbrink\\
\textit{\large Universit\"at M\"unster, Institut f\"ur Theoretische Physik}\\
\textit{\large Wilhelm-Klemm-Str.~9, D-48149 M\"unster, Germany}\\
\textit{\large E-mail: g.bergner@uni-muenster.de}\\[8mm]
I.~Montvay\\
\textit{\large Deutsches Elektronen-Synchrotron DESY}\\
\textit{\large Notkestr. 85, D-22603 Hamburg, Germany}
\vspace*{5mm}}

\date{June 10, 2012 (rev. August 30, 2012)}

\maketitle

\begin{abstract}
The spectrum of particles in supersymmetric Yang-Mills theory is expected to
contain a spin 1/2 bound state of gluons and gluinos, the gluino-glue
particle. We study the mass of this particle in softly broken supersymmetric
Yang-Mills theory on a lattice by means of numerical simulations. The main
focus is the estimation of finite size effects. We extrapolate the mass
first to the infinite volume and then to the limit of a vanishing gluino
mass. The results indicate that finite size effects are tolerable on
lattices of moderate size, and that remaining deviations from supersymmetry
are probably due to finite lattice spacing effects.
\end{abstract}
\section{Introduction}

Supersymmetric Yang-Mills theory (SYM) describes interacting gluons and
their supersymmetric partners, the gluinos. As it is a gauge theory
containing fermionic degrees of freedom, it is in this respect similar to
QCD. An essential difference, however, is that gluinos are Majorana fermions
in the adjoint representation of the gauge group. In Minkowski space the
(on-shell) Lagrangian of SYM is composed out of the gluon fields $A_{\mu}$
and the gluino field $\lambda$, and reads
\begin{equation}
\mathcal{L}=\tr\left[-\frac{1}{4}
F_{\mu\nu}F^{\mu\nu}+\frac{\I}{2}\bar{\lambda}\gamma^\mu
D_\mu\lambda{-\frac{m_g}{2}\bar{\lambda}\lambda} \right ] \,,
\end{equation}
where $F_{\mu\nu}$ is the usual non-Abelian field strength and $D_\mu$
denotes the gauge covariant derivative in the adjoint representation. The
fields $\lambda$ and $A_{\mu}$ are transformed into each other by the
supersymmetry transformation. The gluino mass term breaks supersymmetry
softly.

Some properties of SYM are expected to be similar to QCD
\cite{Amati:1988ft}. It is asymptotically free and is assumed to show
confinement. The determination of its low-energy properties, including the
spectrum of particles, is a non-perturbative problem. The ``physical''
particles are bound states of gluons and gluinos, and if supersymmetry is
unbroken, they would form supermultiplets.

There are several motivations for the numerical investigation of SYM on the
lattice. One of them is to understand the non-perturbative interaction of
the gluinos in supersymmetric extensions of the standard model. There exist
theoretical predictions for the low energy effective theory
\cite{Veneziano:1982ah,Farrar:1997fn} that can be compared to the lattice
results. Another motivation for the numerical investigations of SYM on the
lattice is related to possible connections to ordinary QCD, as provided by
the orientifold planar equivalence \cite{Armoni:2004uu}. Results of previous
investigations of SYM on the lattice by our collaboration can be found in
\cite{Demmouche:2010sf,Bergner:2011wf}.

In recent years, SYM on the lattice has also been investigated with
Ginsparg-Wilson fermions, in the domain wall
\cite{Fleming:2000fa,Giedt:2008xm,Endres:2009yp} as well as in the overlap
formulation \cite{Kim:2011fw}. For large lattice volumes and small lattice
spacings these formulations require, however, a significantly larger amount
of computing resources than the Wilson formulation. The gain of no need for
tuning the position of the zero gluino mass point does not compensate by far
the advantage of Wilson fermions.

Our current studies of SYM are focussed on the bound states of gluons and
gluinos, in particular on an exotic particle state in the spectrum that
arises due to the fermions being in the adjoint representation. This
particle is a spin 1/2 Majorana fermion and is called gluino-glue. It can be
created by operators combining the field strength and the gluino field, the
simplest example being
\begin{equation}
\label{eq:gluinocont}
\tilde{O}_{g\tilde{g}}=
 \sum_{\mu\nu} \sigma_{\mu\nu} \tr \left[ F^{\mu\nu}  \lambda\right],
\end{equation}
with $\sigma_{\mu\nu}=\frac{1}{2} \left[ \gamma_\mu,\gamma_\nu \right]$.
Such a bound state containing a single fermion does not occur in QCD, but
analogous particles exist in models similar to QCD with an arbitrary number
of quark flavours in the adjoint representation.

Since supersymmetry is generically broken in any non-trivial theory on the
lattice \cite{Bergner:2009vg} it has to be ensured that it is restored in
the continuum limit. A necessary condition for unbroken supersymmetry is the
degeneracy of fermionic and bosonic masses. In a supersymmetric theory the
fermionic gluino-glue state must therefore be part of a multiplet containing
also bosonic particles with the same mass. The behaviour of the gluino-glue
particle provides an important signal for the supersymmetric limit of the
theory.

It is expected that in SYM a fine-tuning of the bare gluino mass parameter
in the continuum limit is enough to approach the symmetries of the continuum
theory \cite{Curci:1986sm}. These symmetries include chiral symmetry and
supersymmetry. The theoretical prediction of the existence of a
supersymmetric chiral continuum limit needs to be confronted with the
numerical lattice simulations. The chiral limit, the continuum limit, and
the infinite volume limit can only be extrapolated from the results of these
simulations. In QCD there are good estimates about the scales and parameters
that are necessary to get reliable results, together with systematic
estimates of the induced errors. In SYM the influence of parameters like the
finite volume and non-zero lattice spacing may be different. It is therefore
important to have a detailed understanding of their effects in order to be
able to approach the supersymmetric limit.

The previous results of our simulations have not yet shown the expected
degeneracy of the fermionic and bosonic masses
\cite{Demmouche:2010sf,Bergner:2011wf}. The obtained mass of the gluino-glue
has been larger than the other masses of its lightest possible
superpartners. However, the masses were obtained at a fixed lattice spacing
and without a detailed analysis of the finite size effects. Our most recent
investigations indicate that the influence of the finite lattice spacing is
larger than expected. This provides a possible source of the supersymmetry
breaking in the simulation.

In this work we want to scrutinize the influence of the finite volume. If it
is large, it provides another possible explanation for the gap between
fermionic and bosonic masses. If it is smaller than expected then the
lattice artifacts have to be reduced by performing simulations at smaller
lattice spacings.

In our studies the mass of the gluino-glue particle has typically been the
one determined with the best accuracy. Therefore it is well suited to
estimate the effects of the finite lattice size. In order to estimate the
influence on supersymmetry breaking, the mass of the adjoint version of the
$\eta'$ meson ($\aetap$, a pseudoscalar bound state of gluinos) is also
considered.

\section{Supersymmetric Yang-Mills theory on the lattice}

In our investigations we have employed the lattice formulation of SYM
proposed by Curci and Veneziano \cite{Curci:1986sm}. The gauge field is
represented by link variables $U_\mu(x)$ in the gauge group SU($N_c$). The
corresponding gauge action is the Wilson action built from the plaquette
variables $U_p$. The gluinos are described by Wilson fermions in the adjoint
representation. In its basic form the complete lattice action reads
\begin{equation}
  \mathcal{S}_L=\beta \sum_p\left(1-\frac{1}{N_c}\mbox{Re}\,\tr U_p\right)
   +\frac{1}{2}\sum_{xy} \bar{\lambda}_x(\Di)_{xy}\lambda_y\, ,
\end{equation}
where $\Di$ is the Wilson-Dirac operator
\begin{eqnarray}
 (\Di)_{x,a,\alpha;y,b,\beta}
    &=&\delta_{xy}\delta_{a,b}\delta_{\alpha,\beta}\nonumber\\
    &&-\kappa\sum_{\mu=1}^{4}
      \left[(1-\gamma_\mu)_{\alpha,\beta}(V_\mu(x))_{ab}
                          \delta_{x+\mu,y} \right .\nonumber \\
    &&\left. +(1+\gamma_\mu)_{\alpha,\beta}(V^\dag_\mu(x-\mu))_{ab}
                          \delta_{x-\mu,y}\right].
\end{eqnarray}
The hopping parameter $\kappa$ is related to the bare gluino mass via
$\kappa=1/(2m_g+8)$.

While the link variables $U_{\mu}(x)$ are in the fundamental representation
of the gauge group, the variables $V_{\mu}(x)$ in the Wilson-Dirac operator
are the corresponding elements in the adjoint representation. In our present
investigations the gauge group is SU(2), the fundamental representation is a
doublet and the adjoint one a triplet. In this case the adjoint gauge field
is given by $[V_\mu(x)]^{ab}=2\,\tr [ U^\dag_\mu(x) T^a U_\mu(x) T^b ]$,
where $T^a$ are the generators of the gauge group, normalised such that
$2\,\tr [T^a T^b]=\delta^{ab}$.

In our simulations the basic form of the lattice action has been modified
with the following improvements: in order to reduce the lattice artifacts a
tree-level Symanzik improved gauge action has been used instead of the
simple Wilson gauge action. One level of stout smearing has been applied to
the link fields in the Wilson-Dirac operator to reduce lattice artifacts
also in the fermionic part of the action.

The lattice action explicitly breaks supersymmetry and in addition the
chiral U(1)$_R$ symmetry. However, to recover the continuum symmetries the
necessary fine-tuning of supersymmetry and the U(1)$_R$ symmetry can be
achieved through the same parameter, the bare gluino mass (i.e.\ $\kappa$)
\cite{Curci:1986sm}. The supersymmetric continuum limit coincides with the
chiral continuum limit of the theory.

An important similarity of SYM to Yang-Mills theory is confinement. The
static potential between quark sources in the fundamental representation of
the gauge group shows no signal of a string breaking. The low energy
effective action is built out of bound states of the elementary fields. The
glueballs of Yang-Mills theory are completed with the mesonic states
(gluino-balls) and the fermionic gluino-glue particle. The proposed
supersymmetric low energy effective actions are constructed from two
multiplets of particles \cite{Veneziano:1982ah,Farrar:1997fn}. The lighter
multiplet consists of glueballs and a gluino-glue state. The heavier
multiplet is built from the mesons and a gluino-glue state. As required by
supersymmetry, each multiplet consists of a scalar, a pseudoscalar, and a
fermionic spin 1/2 particle.

On the lattice the $0^{++}$ and the $0^{-+}$ glueball operators are
constructed from the link variables. The mesonic particles are the $\afn$
represented by the operator $\bar{\lambda}\lambda$ and the $\aetap$ by   
$\bar{\lambda} \gamma_5 \lambda$. Since the model contains only one      
species (``flavour'') of gluinos, all possible mesons are ``flavour      
diagonal''. Consequently, their correlation functions contain disconnected
parts in addition to the connected ones, as it is the case in QCD for 
flavour diagonal mesons.

In addition to the above mesons, we consider the adjoint pion ($\api$),
which is the pion in the corresponding theory with two Majorana fermions
in the adjoint representation. The correlator of this particle is the   
connected contribution of the $\aetap$ correlator. The $\api$ is not a  
physical particle in SYM, which only contains one Majorana fermion.     
However, it can be defined in a partially quenched setup, in which the  
model is supplemented by a second species of gluinos and the corresponding
bosonic ghost gluinos, in the same way as for one-flavour QCD 
\cite{Farchioni:2007dw}.

The reason for considering the $\api$ is the following. In the limit of a
vanishing gluino mass, the bare parameter $\kappa$ has to be tuned to the
critical value $\kappa_c$ (chiral limit) that corresponds to the chiral  
theory in the continuum limit. The value of $\kappa_c$ is most easily    
obtained from the dependence of the $\api$-mass on $\kappa$. On the basis
of arguments involving the OZI-approximation of SYM 
\cite{Veneziano:1982ah}, the adjoint pion mass is expected to vanish for a
massless gluino. The $\api$ yields a more precise signal for the tuning   
than the supersymmetric Ward identities. However, in previous studies we  
have checked that both signals lead to a consistent value of $\kappa_c$   
\cite{Demmouche:2010sf}.

For both the $\afn$ and $\aetap$ mesons the correlators contain a
significant contribution from disconnected diagrams (see
\secc{sec:techmass}), especially on larger lattices and small adjoint pion
masses. As in comparable cases in lattice QCD, this contribution leads to a
bad signal to noise ratio in the correlators. Using the techniques detailed
in \secc{sec:techmass}, the gluino-glue mass can be obtained with an
accuracy significantly better compared to the other particles of the
spectrum (apart from the unphysical adjoint pion).

In previous investigations the predicted multiplet structure of the
particles has not been found in the mass spectrum \cite{Demmouche:2010sf}.
The mass of the gluino-glue particle appeared to be heavier than the masses
of the scalar and pseudoscalar states. A breaking of supersymmetry in the
theory on the lattice is induced by the finite lattice spacing and the
finite lattice extent. The former is an unavoidable consequence of the
discretisation; the latter is due to the anti-periodic (thermal) boundary
conditions implemented for the gluinos. It is important to have an estimate
of these effects. If they had an opposite sign, they could compensate each
other and suggest a wrong supersymmetric point. Alternatively, they could
sum up to the total supersymmetry breaking and it would be necessary to
investigate both of them in order to disentangle their effects.

In our numerical simulations we have applied a polynomial hybrid Monte Carlo
(PHMC) algorithm. In a two step procedure, in the Metropolis step a better
polynomial approximation is used than in the integrator of the molecular
trajectory. The remaining error is compensated by a reweighting. Further
details of the simulations algorithm can be found in
\cite{Montvay:2005tj,Demmouche:2010sf}.

\begin{figure}[!htb]
\begin{center}
\includegraphics{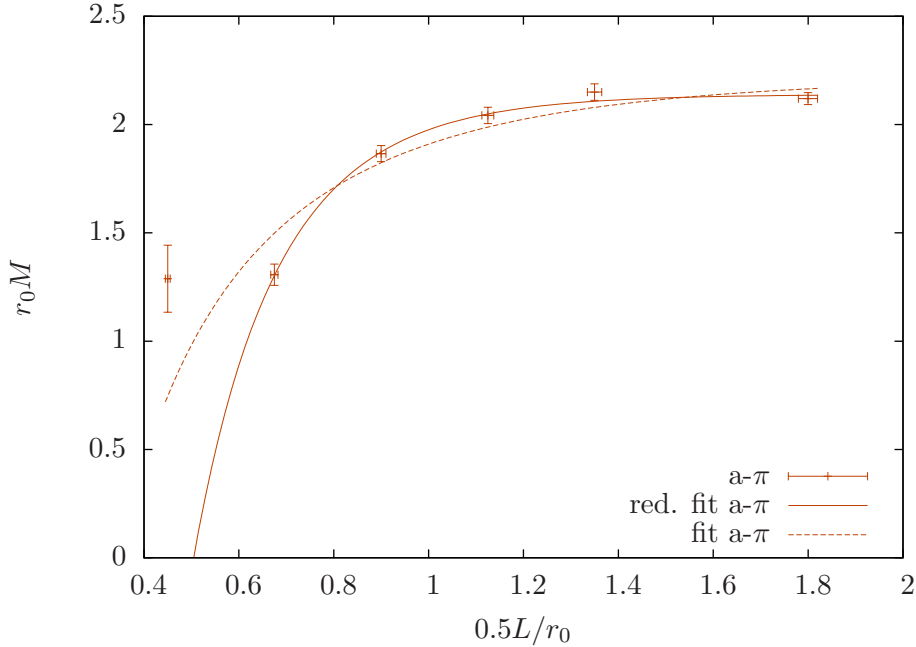}
\caption{The mass of the adjoint pion ($\api$) at different lattice volumes
for a fixed value of $\kappa=0.1490$. The fit (fit) is done using the ansatz
\eqq{eq:fitansatz}, and the reduced fit (red.\ fit) discards the smallest
lattice volume. $r_0$ is the Sommer scale parameter.}
\label{fig:fixedkappapion}
\end{center}
\end{figure}

The simulation at the smallest lattice spacings requires a large number of
configurations to get a reasonable statistical error of the correlation
functions. A further problem arises from the dependence of the adjoint pion
mass on the lattice volume. This dependence is shown in
\figg{fig:fixedkappapion} for a fixed value of $\kappa=0.1490$. The
deviation of the smallest lattice from the infinite volume limit is most
likely due to a larger influence of the excited states in the determination
of the mass.

The adjoint pion mass gets smaller on the smaller lattices. Hence, for a
fixed value of $\kappa$ the simulations are shifted towards the chiral limit
when the lattice volume is reduced. In the vicinity of the chiral limit low
eigenvalues of the Hermitian Wilson-Dirac operator occur. The polynomial
approximation in the PHMC algorithm has a larger error in this case and a
compensation with correction factors is necessary. To obtain these factors
we have calculated an approximation of the lowest eigenvalue on each of the
configurations. Whenever this value has been below the threshold determined
by the polynomial approximation, the correct fermionic contribution of the
100 lowest eigenvalues has been calculated. From this contribution the
correction factors of the reweighting step have been obtained. This
reweighting leads to an additional increase of the statistical error on the
smaller lattice volumes. A complete summary of the simulations is shown in
\tabb{tab:summary}.

\section{Determination of masses}
\label{sec:techmass}

The operator for the gluino-glue particle is represented by the lattice
version of \eqq{eq:gluinocont}
\begin{equation}
O^\alpha_{g\tilde{g}}= \sum_{i<j, \beta} \sigma^{\alpha\beta}_{ij}\,
\tr \left[ P_{ij} \lambda^\beta \right] \, ,
\end{equation}
where the indices $i$ and $j$ stand for the spatial directions. In order to
have the same properties with respect to parity transformation and time
reversal, $F_{\mu\nu}$ is represented by the anti-Hermitian part of the
clover plaquette $U^{(c)}$ \cite{Donini:1997hh},
\begin{equation}
 P_{ij}=\frac{1}{8 \I g_0} (U^{(c)}_{\mu\nu}-(U^{(c)}_{\mu\nu})^\dag)\, .
\end{equation}
The clover plaquette is a combination of links in the fundamental
representation,
\begin{eqnarray}
U^{(c)}_{\mu\nu}&=& U_\mu(x)U_\nu(x+\mu)U_\mu^\dag(x+\nu) U_\nu^\dag(x) \nonumber\\
&& +U^\dag_\nu(x-\nu)U_\mu(x-\nu)U_\nu(x-\nu+\mu) U^\dag_\mu(x)\nonumber\\
&& +U_\mu^\dag(x-\mu)U_\nu^\dag(x-\mu-\nu)U_\mu(x-\mu-\nu) U_\nu(x-\nu)\nonumber\\
&& +U_\nu(x)U_\mu^\dag(x+\nu-\mu)U_\nu^\dag(x-\mu) U_\mu(x-\mu) \nonumber \, .
\end{eqnarray}

The best signal is obtained from the contribution to the correlator that is
proportional to the identity in Dirac space. The corresponding correlator is
\begin{eqnarray}
C_{g\tilde{g}}(x_0-y_0)&=&-\frac{1}{4} \sum_{i,j,k,l}\sum_{\vec{x},\vec{y}}
                                \sum_{\alpha,\beta,\rho}\sum_{ab}\nonumber \\
 &&\langle \sigma_{ij}^{\alpha \beta}
       \tr  [ P_{ij}(x) T^a ] (\Di^{-1})_{x,a,\beta,y,b,\rho}
       \tr [ P_{kl}(y)T^b  ]\sigma_{kl}^{\alpha \rho} \rangle \, .
\end{eqnarray}

At large distances the correlator has the functional form
\begin{equation}
 \label{eq:massfit}
 C_{g\tilde{g}}(t)\approx C \sinh(m (t-T/2))\, ,
\end{equation}
where $T$ is the temporal extent of the lattice. The mass $m$ of the
particle can be obtained by fitting the correlator to this function. The
appropriate $t$-range for the fit, where $t$ is large enough but still much
smaller than $T$, can be found by plotting the effective mass
$m_\text{eff}(t)$. $m_\text{eff}(t)$ is the parameter $m$ obtained from the
correlator at $t$ and $t+1$ assuming \eqq{eq:massfit} to be valid. In the
$t$-range, where the influence of the excited states is small enough,
$m_\text{eff}(t)$ shows a plateau. In a more refined approach the correlated
chi-square of the fit provides an indication that the assumption of the
\eqq{eq:massfit} is a good approximation in the considered region. We have
applied a method that combines fit values of several fitting intervals and
their correlated chi-square to get an estimation of the mass and the
statistical and systematic error \cite{Baron:2010th,Montvay:2012zz}.

The gluino-glue correlator has been obtained using different smearing
techniques. The link fields are smeared using APE smearing, the fermionic
fields using Jacobi smearing. Without any smearing the links $U$ in
$P_{\mu\nu}$ should be the same fundamental links as in the gauge part of
the action. The Wilson-Dirac operator $\Di$, on the other hand, should
contain the one-level stout smeared links in the adjoint representation,
that are used in the fermionic part of the action. In addition to this
approach, the one-level stout smearing has also been applied to the $U$
fields. It has been checked that the smearing in the time-like direction
included in this step has no influence on the correlation function at
distances relevant for the mass determination.

\begin{figure}[!htb]
\begin{center}
\includegraphics{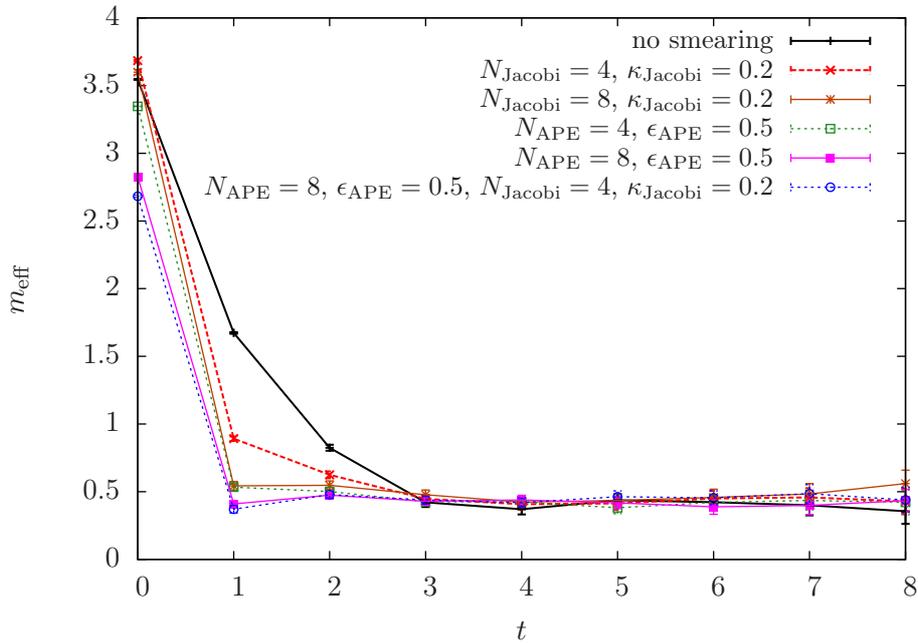}
\caption{The effective mass of the gluino-glue on a $16^3\times 36$ lattice
and $\kappa=0.1492$. The impact of different smearing parameters are
compared. APE smearing is used on the the link fields in $P$, Jacobi
smearing is applied on the fermionic source and sink.}
\label{fig:smearedgluino}
\end{center}
\end{figure}

The effect of the smearing on the effective masses is shown in
\figg{fig:smearedgluino}. The same Jacobi smearing is applied on the source
and sink side of the correlator. The APE and the Jacobi smearing have a
similar impact on the correlator. The excited states at small distances are
considerably reduced by the smearing procedure.

The correlator for the $\aetap$ boson contains a connected and a
disconnected contribution,
\begin{eqnarray}
 C_{\saetap}&=&C_{\sapi}-C_{\saetap \text{ disc}}\nonumber \\
            &=& \frac{1}{L^3} \sum_{\vec{x}}\langle\tr \left[ \gamma_5 (\Di)^{-1}_{x,y}
                  \gamma_5 (\Di)^{-1}_{y,x} \right] \rangle \nonumber \\
            && - \frac{1}{2 L^3} \sum_{\vec{x}} \langle \tr \left[
                   \gamma_5 (\Di)^{-1}_{x,x} \right] \tr \left[ \gamma_5 (\Di)^{-1}_{y,y}
                       \right]\rangle\, .
\end{eqnarray}
Its connected part is the correlator of the adjoint pion ($C_{\sapi}$). The
disconnected part has been calculated using the stochastic estimator method
\cite{Bali:2009hu}. The statistical fluctuations in this part are usually
larger than in the connected contribution.

\begin{figure}[!htb]
\begin{center}
\includegraphics{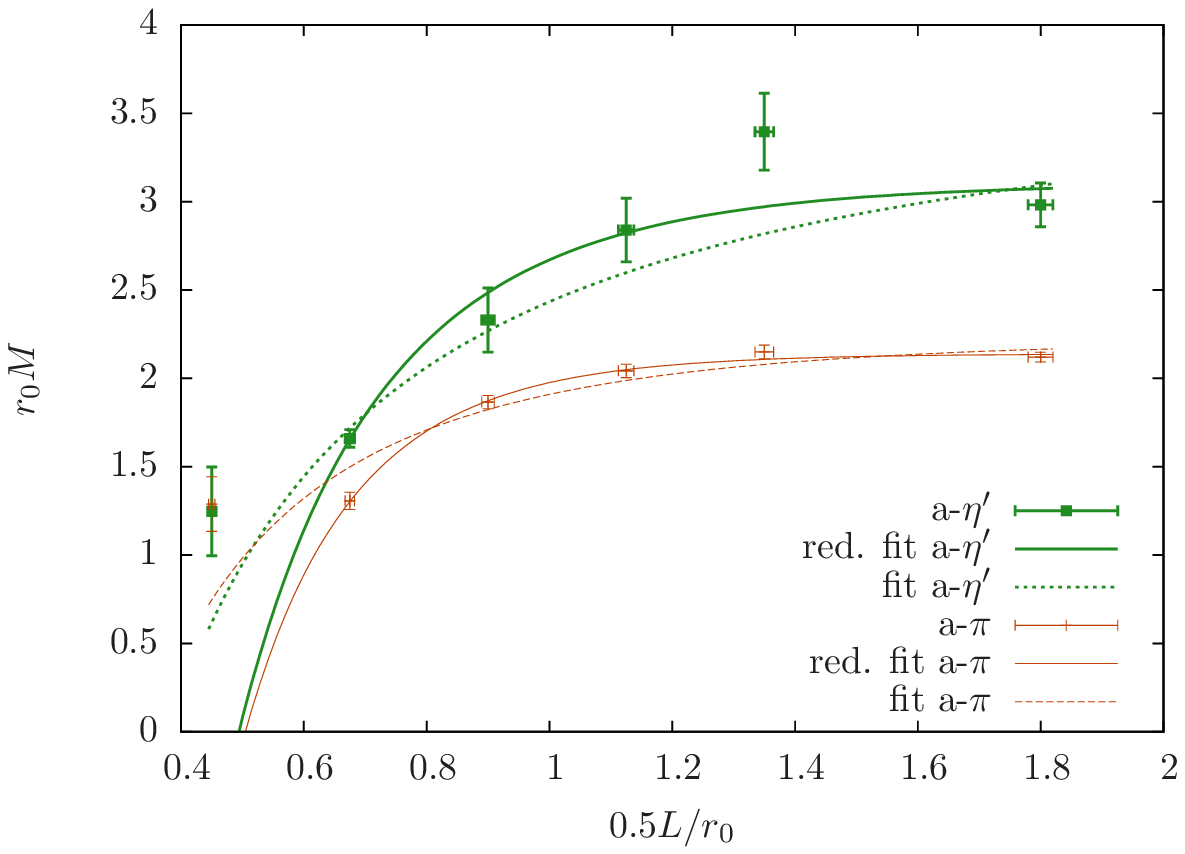}
\caption{The masses of the $\aetap$ and the adjoint pion ($\api$) at
different lattice volumes for a fixed value of $\kappa=0.1490$. The fit (fit) is
done using the ansatz \eqq{eq:fitansatz}, and the reduced fit (red.\
fit) discards the smallest lattice volume.}
\label{fig:fixedkappaetapion}
\end{center}
\end{figure}

It is instructive to compare the masses of the $\aetap$ and the $\api$,
since their difference shows the influence of the disconnected part. In
\figg{fig:fixedkappaetapion} the dependence of these masses on the lattice
volume is shown. On the smallest lattice both masses deviate from the
behaviour expected from the fits. This is presumably due to the larger
contribution of excited states to the correlator, since only smaller values
of $t$ enter. A second observation is that at smaller lattice volumes both
masses appear to approach each other. Thus at smaller volumes the influence
of the disconnected part of the $\aetap$ correlator is relatively smaller
compared to the connected one.

At small $t$ the contribution of the connected part is much larger than the
disconnected part, whereas at intermediate distances and close to the chiral
limit the disconnected part yields a significant contribution. The larger
statistical error of the $\aetap$ on the largest lattices is due to the
larger statistical fluctuations of the disconnected part. The total error on
a small lattice seems to be underestimated for the $\aetap$ particle due to
these systematic uncertainties. Smearing techniques or a larger temporal
extent of the lattice could lead to better results. Because of these
systematic uncertainties of the $\aetap$ mass and the larger statistical
error, the best signal for the finite size effects is obtained from the
gluino-glue mass.

\section{Finite size effects}

In quantum field theory in a finite volume the propagation and interactions
of particles are different from those in an infinite volume. This can be
understood as the modification of the polarized vacuum within a Compton
wavelength around a particle due to the deformation by the finite boundaries
of the box. The infinite volume mass $m_0$ of the particle is shifted due to
the influence of the finite box size $L$,
\begin{equation}
m(L)=m_0+\Delta m(L)\, .
\end{equation}

The mass shift $\Delta m(L)$ has been studied to all orders in perturbation
theory in massive field theories in the continuum in \cite{Luscher:1985dn}.
A similar investigation in the framework of field theory on the lattice has
been made in \cite{Munster:1984zf}. The asymptotic behaviour was found to be
\begin{equation}
\label{eq:fitansatz}
\Delta m(L) \approx C  L^{-1} \exp \left( -\alpha m_0L\right),
\end{equation}
with parameters $C$ and $1 \leq \alpha \leq \sqrt{3}/2$. This behaviour is
quite generic, since it does not depend on the specific form of the
interactions. It also applies to the masses of stable bound states in a
confining theory. For glueballs in lattice gauge theory the constants are
\cite{Munster:1984zf}
\begin{equation}
 C=-\frac{3}{16 \pi} \frac{\lambda^2}{m_0^2}\,,\quad \alpha=\frac{\sqrt{3}}{2}\,,
\end{equation}
where $\lambda$ is the three-glueball coupling constant. We fitted the
dependence of the masses on the finite box size $L$ by the general
asymptotic relation \eqq{eq:fitansatz} and obtained in this way an
extrapolation to the infinite volume limit. To get more stable results, we
have used a numerical fit to obtain the constants $m_0$, $C$, and $\alpha$.

We have carried out simulations at several box sizes $L$, see
\tabb{tab:summary}. The temporal extent $T$ of the lattice has been chosen
to be about twice the spatial extent $L$. This combination has been chosen
since it is the usual setup in all our simulations.

\begin{figure}[!htb]
\begin{center}
\includegraphics{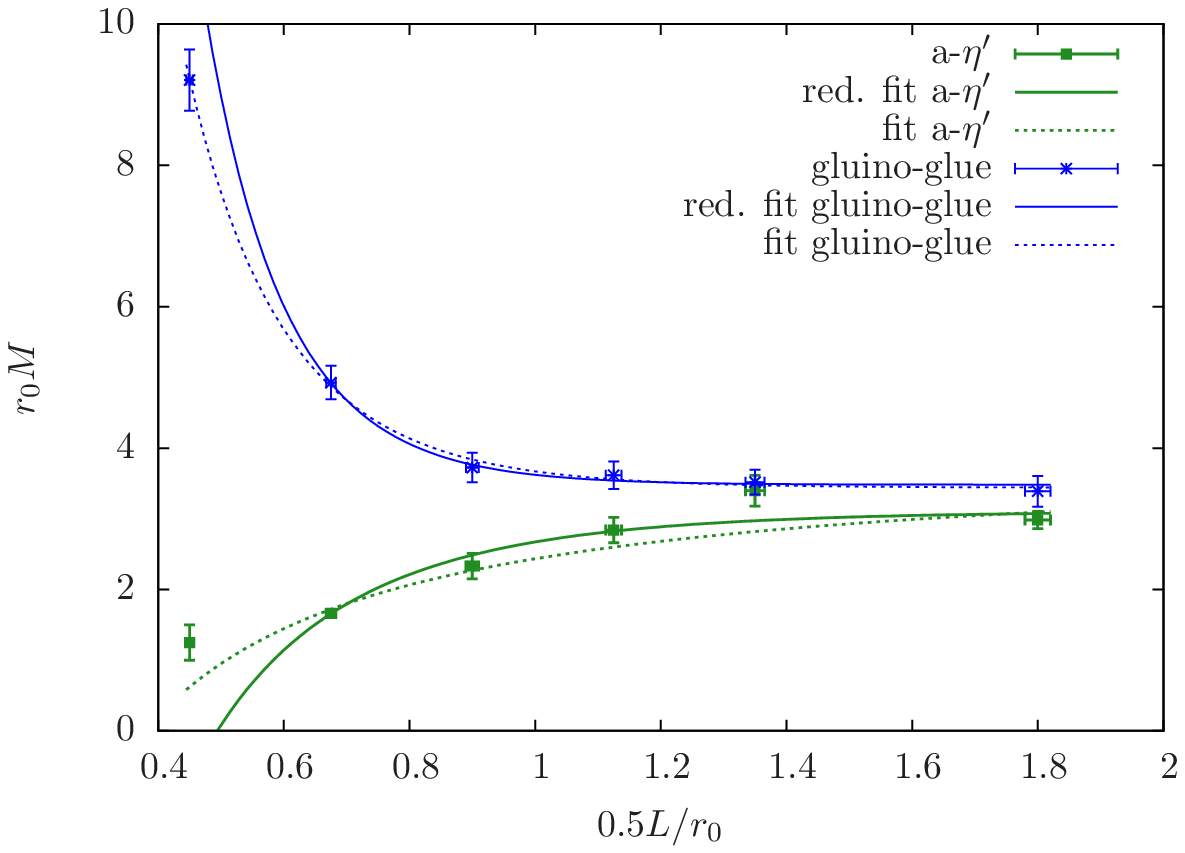}
\caption{The masses of the gluino-glue and the $\aetap$ meson at different
lattice volumes for a fixed value of $\kappa=0.1490$. The fit (fit) is done
using the ansatz \eqq{eq:fitansatz}, and the reduced fit (red.\ fit)
discards the smallest lattice volume.}
\label{fig:fixedkappa}
\end{center}
\end{figure}
The first estimation of finite size effects is done at a fixed value of the
bare gluino mass defined by the hopping parameter $\kappa=0.1490$. To
visualize the influence on the mass gap between bosonic and fermionic
states, the gluino-glue mass is shown in \figg{fig:fixedkappa} together with
the mass of the bosonic $\aetap$ meson. Our ansatz \eqq{eq:fitansatz} for
the functional dependence of the mass on the lattice volume is valid only in
the asymptotic region of large $L$ and might fail for the smallest lattice
sizes. To check its validity we have done a second fit (red.\ fit) that
excludes the smallest lattice volume. The dimensionless scale $0.5 L/r_0$
corresponds to the length in femtometers if the Sommer parameter $r_0$ is
set to $r_0=0.5 \,\mathrm{fm}$ as in QCD. Note that we have always taken the
value of $r_0$ obtained by an extrapolation to the chiral limit of the data
obtained on the $24^3\times 48$ and $32^3\times 64$ lattices.

The gluino-glue gets a positive mass shift at smaller volumes, whereas the
$\aetap$ gets a negative one. This clearly shows that the finite lattice
size enhances the supersymmetry breaking induced by lattice discretisation
and nonvanishing gluino mass in our model: at smaller lattices an additional
splitting of the bosonic and fermionic masses is introduced.

In our previous simulations at larger box sizes the gluino-glue particle has
been the heaviest of the measured low energy states. Since finite size
effects are small in this case, this indicates that the mass shift induced
by the discretisation artifacts is also positive. So, both finite size and
finite lattice spacing effects add up to a total mass splitting between
bosonic and fermionic states indicating the breaking of supersymmetry.

In case of the gluino-glue, \eqq{eq:fitansatz} seems to describe the volume
dependence of the mass better than for the $\aetap$. Also, the fit for the
$\aetap$ fails for the smallest lattice volume. One possible reason for the
different behaviour is the smearing of the gluino-glue that reduces the
contribution of the excited states. This reduction becomes important for the
smallest lattice sizes, where the masses are obtained from fits of the
correlation function at smaller distances. Another reason are the systematic
uncertainties in the mass determination of the $\aetap$ at smaller lattice
sizes (see \secc{sec:techmass}). Nevertheless, neglecting the smallest
lattice size, the fit seems to reproduce the general behaviour of the mass
gap at different lattice volumes.

Above a box size of about $L=1.2\, r_0/0.5$ (corresponding to 1.2 fm in QCD
units) the statistical errors and the systematic errors of the finite size
effects are of the same order and hence the finite size effects can be
neglected at present. This is an important estimate for the minimal lattice
size required for the simulations.

In the following we focus our discussion on the gluino-glue since the signal
of the finite size effects is much clearer for this particle. Final results
for masses in SYM should be obtained as extrapolations towards the infinite
volume limit, the continuum limit, and the chiral limit. The continuum limit
will be the subject of future investigations. The closer one gets to the
chiral limit, the smaller are the masses of the particles and the ratio of
the box size to the Compton wavelength. In other words, the finite size
effects are expected to get larger in the vicinity of the chiral limit.
Therefore, the proper procedure is an extrapolation to the infinite volume
limit before the extrapolation of the chiral limit.

\begin{figure}[p]
\begin{center}
\includegraphics{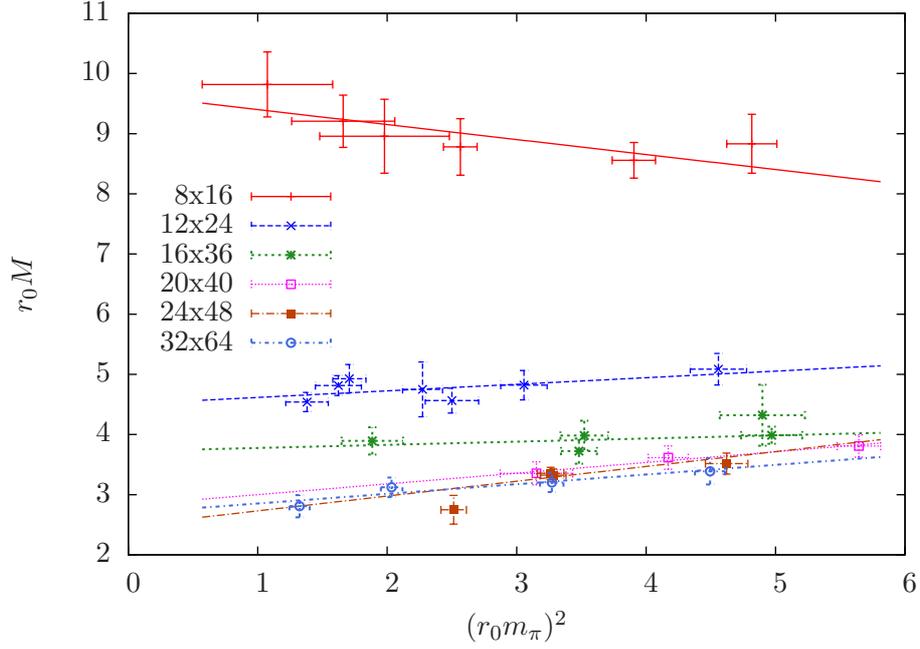}
\caption{The mass of the gluino-glue at different lattice volumes and
different values of $\kappa$ as a function of the square of the adjoint pion
mass in units of the Sommer scale. The lines are obtained from a linear
regression of the points for each lattice volume. They are used as a linear
interpolation.}
\label{fig:gluinoall}
\vspace*{-10mm}
\end{center}
\end{figure}
\begin{figure}[p]
\begin{center}
\includegraphics[width=7.2cm]{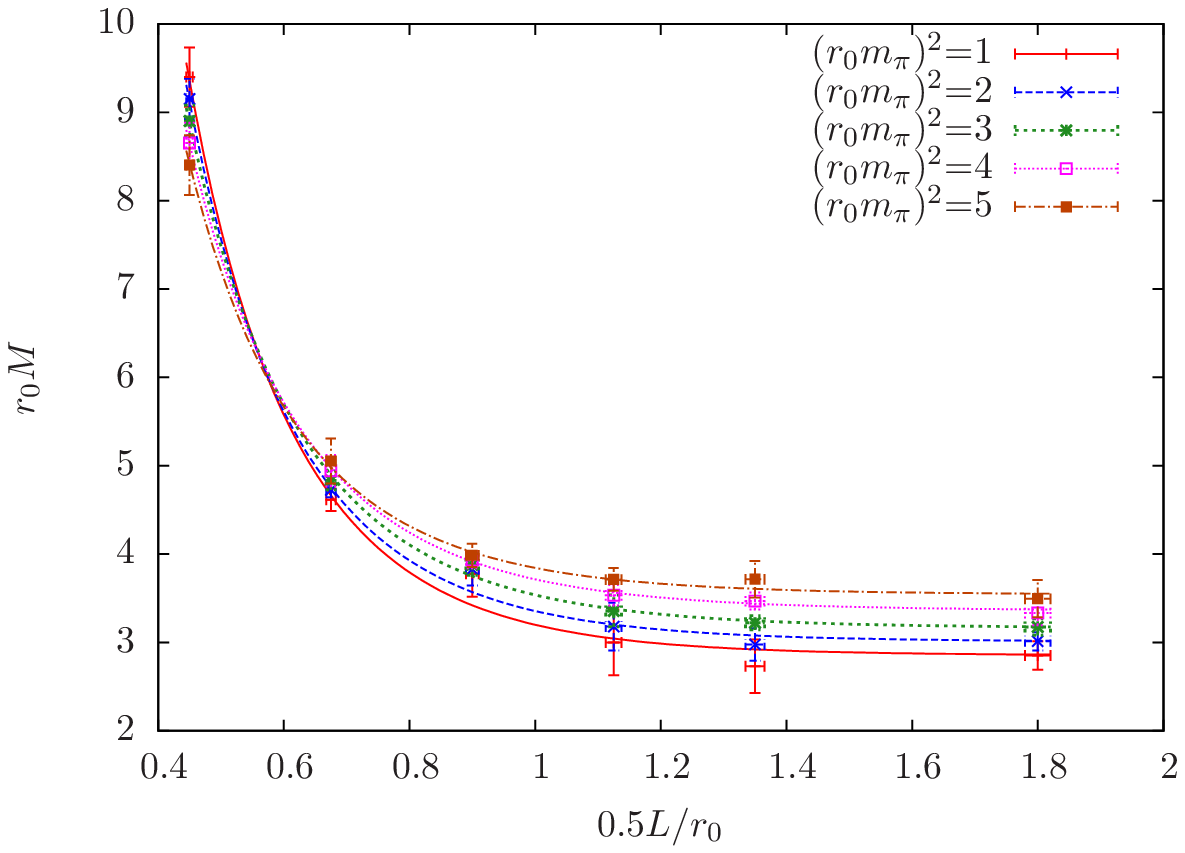}
\includegraphics[width=7.2cm]{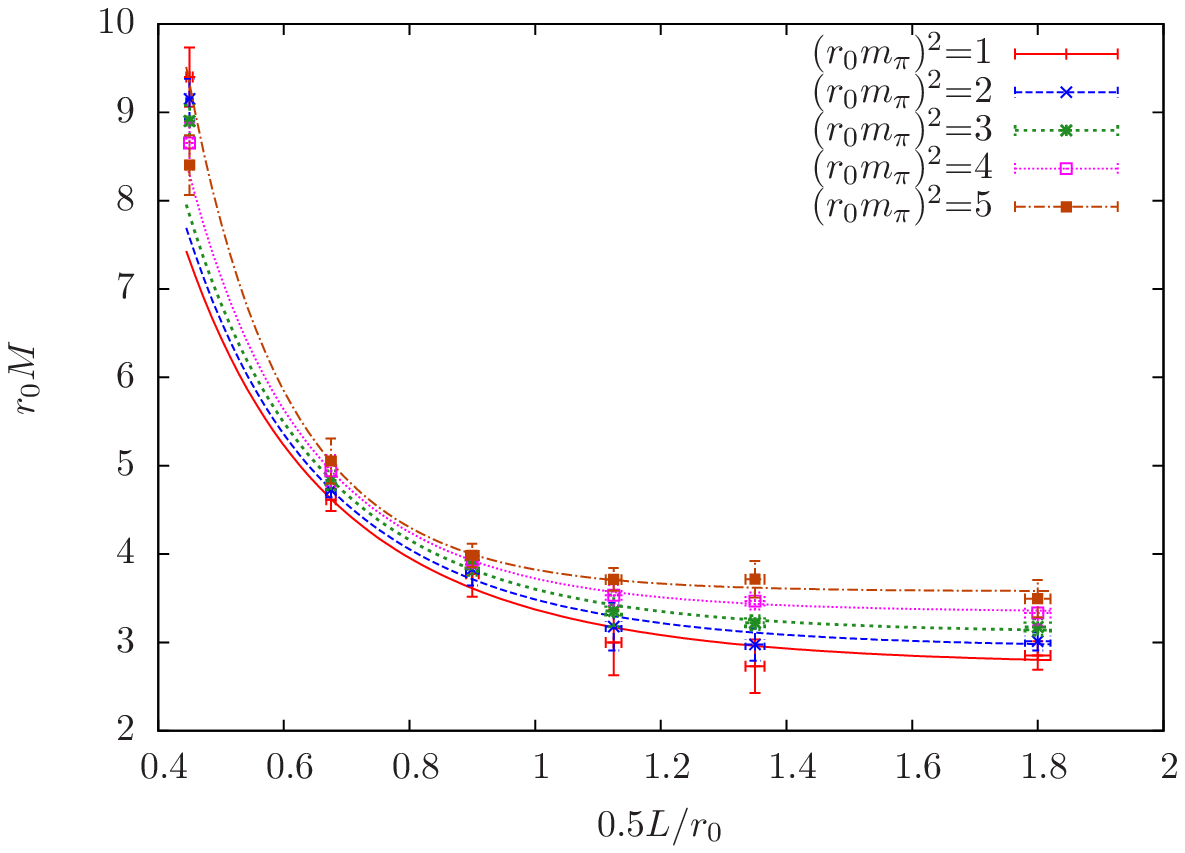}
\caption{A fit of the finite size effects at several fixed values of $(r_0
m_{\sapi})^2$. The fit is done using the ansatz \eqq{eq:fitansatz}. The
right panel shows the fits neglecting the smallest lattice volume.}
\label{fig:gluinoallinter}
\vspace*{-10mm}
\end{center}
\end{figure}
\begin{table}[p]
\begin{center}
\begin{tabular}{|c|c|c|c|c|c|c|}
\hline
$(r_0 m_{\sapi})^2$ & $m_{0r}$ (fit) & $C$ (fit) & $\alpha$ (fit)
  & $m_{0r}$ (red.\ fit) & $C$ (red.\ fit) & $\alpha$ (red.\ fit) \\
\hline
$1$ & $2.85(13)$ & $33.6(71)$ & $0.68(11)$ & $2.74(19)$ & $10.8(84)$ & $0.39(26)$ \\
$2$ & $3.010(81)$ & $30.0(43)$ & $0.627(65)$ & $2.94(11)$ & $12.4(72)$ & $0.41(17)$ \\
$3$ & $3.165(48)$ & $25.0(28)$ & $0.555(46)$ & $3.108(60)$ & $14.0(47)$ & $0.426(92)$ \\
$4$ & $3.362(28)$ & $22.7(16)$ & $0.516(26)$ & $3.345(41)$ & $18.9(52)$ & $0.481(66)$ \\
$5$ & $3.543(56)$ & $22.3(33)$ & $0.510(54)$ & $3.578(70)$ & $35(23)$ & $0.60(15)$ \\
\hline
\end{tabular}
\caption{Fit parameters for the fits of the finite size effects shown in
\figg{fig:gluinoallinter}.}
\label{tab:volumeextr}
\end{center}
\vspace*{-20mm}
\end{table}

In the first step we have extrapolated the values of the masses to the
infinite volume limit but away from the chiral limit. Instead of a fixed
value of the bare parameter $\kappa$, we have fixed the squared mass of the
adjoint pion, $(r_0 m_{\sapi})^2$, to five different values. The masses of
the gluino-glue at these values have been obtained from a linear
interpolation of the simulation results at each box size. These
interpolations are shown in \figg{fig:gluinoall}. The masses at the three
largest box sizes are almost indistinguishable in view of the current
statistical errors.

The results of the interpolation are shown in \figg{fig:gluinoallinter}. The
lines correspond to the fit of the dependence on the finite box size
according to \eqq{eq:fitansatz}. The parameter $m_0$ of the fit is the
extrapolated infinite volume limit of the masses. The different parameters
of the fits are summarised in \tabb{tab:volumeextr}.

\begin{figure}[!htb]
\begin{center}
\includegraphics{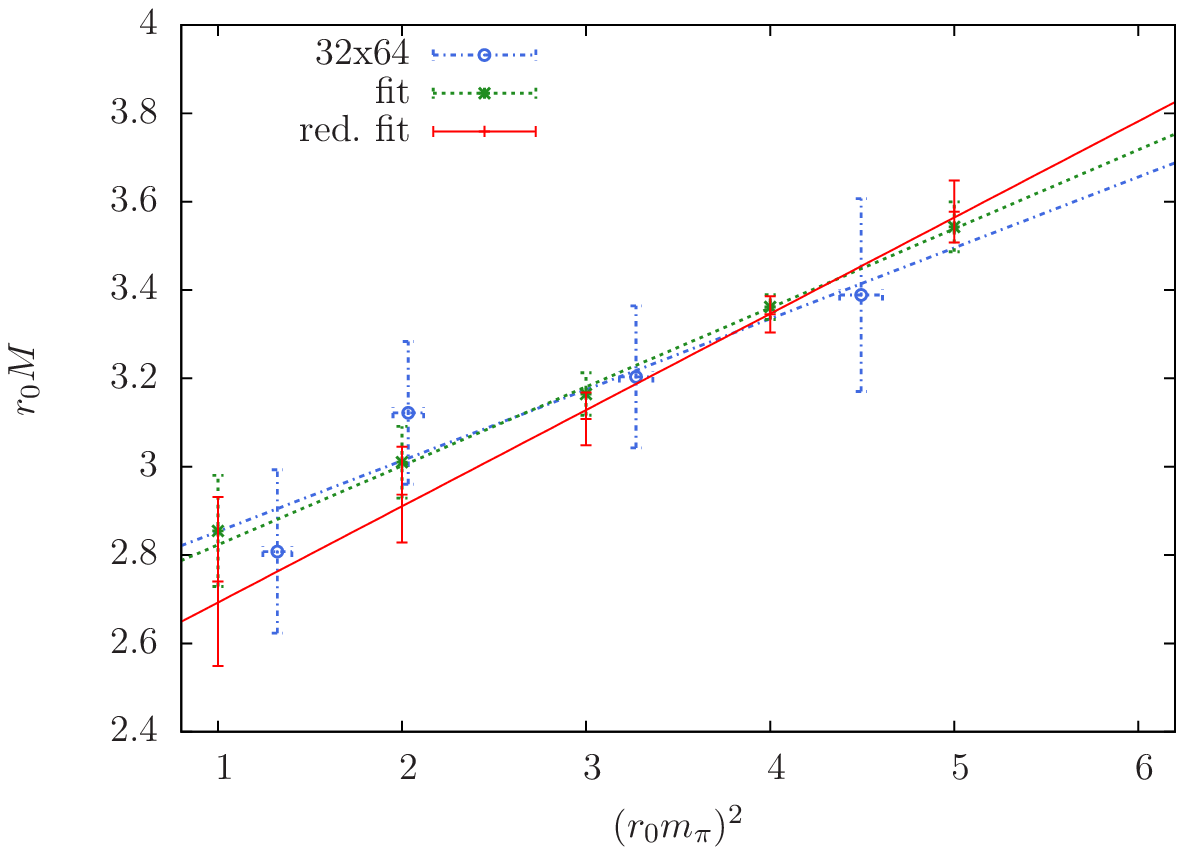}
\caption{The gluino-glue mass extrapolated to the infinite volume limit as a
function of the squared mass of the adjoint pion in units of the Sommer
scale. In addition the mass is shown at the largest lattice volume and
different values of $\kappa$.}
\label{fig:gluinoallinterres}
\end{center}
\end{figure}

The masses extrapolated to the infinite volume are shown in
\figg{fig:gluinoallinterres} as a function of $(r_0 m_{\sapi})^2$. For
comparison, the results of the largest lattice have been added to this plot.

From the masses at different values of $(r_0 m_{\sapi})^2$ the chiral
extrapolation can be obtained assuming a linear dependence. The linear
regression of this dependence can be seen in \figg{fig:gluinoallinterres}.
The values obtained from linear extrapolations to the chiral limit are
displayed in \tabb{tab:chiralextr}. The table contains the values at fixed
lattice sizes as well as the results from the infinite volume
extrapolations. The results at the largest two lattices appear to be
consistent with those from the infinite volume limit.
\begin{table}[H]
\begin{center}
\begin{tabular}{|c|c|c|c|c|c|c|c|}
\hline
$8\times 16$ & $12\times 24$ & $16\times 36$ &  $20\times 40$ & $24\times 48$ & $32\times 64$ & fit         & red.\ fit \\
\hline
 $9.65(46)$  &  $4.51(20)$   & $3.72(34)$    &  $2.82(47)$    & $2.48(42)$    & $2.69(23)$    & $2.644(91)$ & $2.47(12)$ \\
\hline
\end{tabular}
\caption{The lower entries of this table are the results of chiral
extrapolations of the gluino-glue mass at fixed lattice sizes. The last two
entries are obtained by first performing the extrapolation to the infinite
volume limit and then to the chiral limit. In the reduced fit (red.\ fit)
the smallest lattice volume is ignored. All masses are in units of the
Sommer scale.}
\label{tab:chiralextr}
\end{center}
\end{table}

\FloatBarrier
\section{Conclusions}

In this work we have investigated the implications of the finite lattice
extent in the simulations of SYM on a lattice. An exotic particle of SYM,
the gluino-glue, turned out to provide the best signal for the finite volume
dependence of the masses. A comparison with the bosonic $\aetap$ meson shows
that the finite size effects on the masses have opposite signs for both
particles: the gluino-glue gets a larger mass at smaller box size, whereas
the mass of the $\aetap$ is reduced. This indicates that finite size effects
enhance the supersymmetry breaking induced by lattice discretisation and
nonvanishing gluino mass, and increase the unexpected gap between the
bosonic and fermionic states seen in previous lattice results. The influence
of the finite box size decreases, however, quite rapidly. The differences
between the three largest lattices in our simulations are nearly negligible.
The physical reason for this effect is the large mass of the lightest
particle in this theory. The adjoint pion, which is light and becomes
massless in the chiral limit, is not a physical particle and is only defined
in a partially quenched framework. The lightest physical particles have
significantly larger masses, even in the chiral limit. Therefore their
Compton wavelengths are small compared to the sizes of our largest lattices.

The aim of these investigations was to obtain estimates of the finite size
effects and their implications on the setup for future simulations of SYM.
Another item would be to consider the effect of changing the fermionic
temporal boundary conditions from antiperiodic to periodic. This could help
to disentangle the effects of the finite spatial volume and the finite
temporal extent. In our present work we have chosen a ration of $L$ over $T$
similar to our final simulations for the determination of the complete
spectrum on larger lattices.

The findings have implications for our further investigations of SYM. We
have found that finite size effects are small for $L \geq 1.2\, r_0/0.5$
(corresponding to 1.2 fm in QCD units), and that the simulations can be
efficiently done at present lattice spacings on medium lattice sizes, for
instance, $24^3 \times 48$. In order to obtain reliable results relevant to
the continuum limit a sufficiently large statistics and a small lattice
spacing appear to be crucial.

\section*{Acknowledgments}

This project is supported by the German Science Foundation (DFG) under
contract Mu 757/16, and by the John von Neumann Institute for Computing
(NIC) with grants of computing time. Further computing time has been
provided by the compute cluster PALMA of the Unversity of M\"unster.


\pagebreak
\begin{appendix}
\setlength{\textfloatsep}{0cm}
\setlength{\belowcaptionskip}{0cm}
 \section{Summary of the simulations and results}
\nopagebreak
\begin{table}[hb!]
\begin{center}
\begin{tabular}{|c|c|c|c|c|c|c|c|c|}
\hline
$L\times T$ & $0.5L/r_0$ & $\kappa$ & $a m_{\sapi}$ & $(r_0 m_{\sapi})^2$ & $a m_{g\tilde{g}}$ & $a m_{\saetap}$ & $N_\text{conf}$& $N_\text{corr}$\\
\hline
$8\times16$ & $0.4499(51)$ & $0.1475$ & $0.2469(22)$ & $4.82(19)$ & $0.993(44)$ & $0.2388(29)$ &  12573 & -- \\
$8\times16$ & $0.4499(51)$ & $0.1478$ & $0.2223(23)$ & $3.91(17)$ & $0.963(23)$ & $0.2140(30)$ &  12890 & 2 \\
$8\times16$ & $0.4499(51)$ & $0.1482$ & $0.1802(25)$ & $2.56(13)$ & $0.988(42)$ & $0.1760(41)$ &  5062 & 3 \\
$8\times16$ & $0.4499(51)$ & $0.1487$ & $0.158(18)$ & $1.98(50)$ & $1.008(58)$ & $0.159(16)$ &  9900 & 2 \\
$8\times16$ & $0.4499(51)$ & $0.1490$ & $0.145(16)$ & $1.66(40)$ & $1.036(37)$ & $0.140(27)$ &  59789 & 47 \\
$8\times16$ & $0.4499(51)$ & $0.1491$ & $0.117(26)$ & $1.07(50)$ & $1.104(48)$ & $0.108(20)$ &  59972 & 40 \\
\hline
$12\times24$ & $0.6749(76)$ & $0.1485$ & $0.2402(30)$ & $4.56(22)$ & $0.572(23)$ & $0.233(13)$ &  9316 & 1 \\
$12\times24$ & $0.6749(76)$ & $0.1487$ & $0.1967(36)$ & $3.06(18)$ & $0.542(21)$ & $0.2050(43)$ &  9096 & 11 \\
$12\times24$ & $0.6749(76)$ & $0.1488$ & $0.1778(53)$ & $2.50(21)$ & $0.513(18)$ & $0.1711(74)$ &  8774 & 16 \\
$12\times24$ & $0.6749(76)$ & $0.1489$ & $0.1696(39)$ & $2.27(15)$ & $0.534(45)$ & $0.1636(49)$ &  8073 & 20 \\
$12\times24$ & $0.6749(76)$ & $0.1490$ & $0.1470(39)$ & $1.71(13)$ & $0.554(21)$ & $0.1867(35)$ &  8083 & 225 \\
$12\times24$ & $0.6749(76)$ & $0.1491$ & $0.1433(62)$ & $1.62(18)$ & $0.541(12)$ & $0.1604(87)$ &  8238 & 158 \\
$12\times24$ & $0.6749(76)$ & $0.1492$ & $0.1322(64)$ & $1.38(16)$ & $0.511(12)$ & $0.147(12)$ &  8033 & 173 \\
\hline
$16\times36$ & $0.900(10)$ & $0.1487$ & $0.2508(31)$ & $4.97(24)$ & $0.449(11)$ & $0.349(22)$ &  4699 & -- \\
$16\times36$ & $0.900(10)$ & $0.1488$ & $0.2490(56)$ & $4.90(33)$ & $0.486(52)$ & $0.281(14)$ &  1101 & 9 \\
$16\times36$ & $0.900(10)$ & $0.1489$ & $0.2111(31)$ & $3.52(18)$ & $0.448(22)$ & $0.228(32)$ &  4796 & 10 \\
$16\times36$ & $0.900(10)$ & $0.1490$ & $0.2099(18)$ & $3.48(14)$ & $0.419(19)$ & $0.262(17)$ &  9909 & 24 \\
$16\times36$ & $0.900(10)$ & $0.1492$ & $0.1544(80)$ & $1.89(24)$ & $0.438(21)$ & $0.263(17)$ &  7550 & 177 \\
\hline
$20\times40$ & $1.125(13)$ & $0.1488$ & $0.26723(96)$ & $5.64(17)$ & $0.428(16)$ & $0.323(19)$ &  3700 & -- \\
$20\times40$ & $1.125(13)$ & $0.1490$ & $0.2297(17)$ & $4.17(15)$ & $0.407(17)$ & $0.319(17)$ &  3698 & 5 \\
$20\times40$ & $1.125(13)$ & $0.1492$ & $0.1997(66)$ & $3.15(28)$ & $0.377(17)$ & $0.37(11)$ &  3799 & 72 \\
\hline
$24\times48$ & $1.350(15)$ & $0.1490$ & $0.2418(16)$ & $4.62(16)$ & $0.396(15)$ & $0.382(20)$ &  6899 & 11 \\
$24\times48$ & $1.350(15)$ & $0.1492$ & $0.20321(89)$ & $3.26(10)$ & $0.3773(71)$ & $0.301(14)$ &  3228 & 81 \\
$24\times48$ & $1.350(15)$ & $0.1492$ & $0.20381(80)$ & $3.283(99)$ & $0.3740(75)$ & $0.299(28)$ & 8869 & -- \\
$24\times48$ & $1.350(15)$ & $0.1493$ & $0.1783(15)$ & $2.513(98)$ & $0.309(23)$ & $0.289(19)$ &  10203 & 18 \\
\hline
$32\times64$ & $1.800(20)$ & $0.1490$ & $0.23847(41)$ & $4.49(12)$ & $0.381(20)$ & $0.335(10)$ &  5039 & -- \\
$32\times64$ & $1.800(20)$ & $0.1492$ & $0.20346(54)$ & $3.272(91)$ & $0.360(14)$ & $0.292(13)$ &  6348 & -- \\
$32\times64$ & $1.800(20)$ & $0.1494$ & $0.1604(15)$ & $2.034(83)$ & $0.351(14)$ & $0.269(25)$ &  5485 & 156 \\
$32\times64$ & $1.800(20)$ & $0.1495$ & $0.1294(24)$ & $1.323(79)$ & $0.316(17)$ & $0.253(45)$ &  2147 & 80 \\
\hline
\end{tabular}
\caption{Summary of our simulations at $\beta=1.75$. The value of $r_0/a$
obtained from the extrapolation of the $24^3\times 48$ and $32^3\times 64$
lattice data to the chiral limit is $r_0/a=8.89(10)$. $N_\text{conf}$ is the
number of configurations in the gluino-glue mass measurement.
$N_\text{corr}$ is the total number of configurations with a reweighting
factor different from one.}
\label{tab:summary}
\end{center}
\end{table}
\end{appendix}

\end{document}